\pdfoutput=1
\documentclass[aps,prapplied,reprint,superscriptaddress,longbibliography]{revtex4-2}
\usepackage{graphicx}
\usepackage{amsmath, amssymb}
\usepackage{hyperref}
\usepackage{siunitx}
\usepackage{float}
\usepackage{textcomp}
\usepackage{makecell}
\usepackage{listings}
\usepackage{xcolor}

\definecolor{codegray}{rgb}{0.5,0.5,0.5}
\definecolor{codeblue}{rgb}{0.26, 0.45, 0.77}

\lstdefinestyle{mypython}{
	language=Python,
	basicstyle=\ttfamily\small,
	keywordstyle=\color{codeblue}\bfseries,
	commentstyle=\color{codegray}\itshape,
	stringstyle=\color{red},
	showstringspaces=false,
	breaklines=true,
	frame=single,
	numbers=left,
	numberstyle=\tiny\color{codegray},
	captionpos=b
}

\usepackage[utf8]{inputenc}

\begin{document}

\title{Purcell-enhanced lifetime modulation of quantum emitters as a probe of local refractive index changes}

\author{Yevhenii M. Morozov}
\email{yevhenii.morozov@ait.ac.at}
\affiliation{AIT-Austrian Institute of Technology, Vienna, Austria}
\author{Anatoliy S. Lapchuk}
\affiliation{Department of Optical Engineering, Institute for Information Recording of NAS of Ukraine, Kyiv, Ukraine}

\date{August 12, 2025}

\begin{abstract}
Quantum emitters embedded in photonic integrated circuit (PIC) cavities offer a powerful and scalable platform for label-free refractive index sensing at the nanoscale. We propose and theoretically analyze a sensing mechanism based on Purcell-enhanced modulation of the emitter’s spontaneous emission lifetime, enabling detection of refractive index changes via time-correlated single-photon counting (TCSPC). In contrast to traditional resonance-shift sensors, our approach exploits the lifetime sensitivity to variations in the local density of optical states (LDOS), providing an intensity-independent, spectrally unresolvable, and potentially CMOS-compatible sensing modality. We derive analytical expressions linking refractive index perturbations to relative lifetime shifts and identify an optimal off-resonance operation regime where the lifetime response becomes linear and maximally sensitive to small perturbations. As a PIC material, silicon offers seamless integration of single-photon avalanche detectors suitable for TCSPC. We therefore illustrate the applicability of the proposed mechanism for quantum emitters in silicon. Parametric evaluation of the analytical expressions shows that, for moderate-quality photonic cavities ($Q = 10^5$–$10^7$), this method enables refractive index detection limits as low as $10^{-9}$~RIU --- competitive with or even outperforming state-of-the-art plasmonic and microresonator sensors, yet requiring significantly simpler instrumentation. Furthermore, long-lived emitters such as T-centers in silicon offer a unique advantage, allowing sub-nanosecond lifetime shifts to be resolved with standard TCSPC systems. Although room-temperature operation of quantum emitters in silicon has yet to be demonstrated, our results lay the theoretical foundation for scalable, room-temperature, quantum-enabled refractive index sensing --- eliminating the need for spectral resolution and cryogenic infrastructure. Given the generic nature of the proposed approach, and with ongoing advances in PIC technologies such as diamond-on-silicon and hybrid silicon/silicon nitride and silicon/silicon carbide platforms, the method can be readily extended to materials in which room-temperature operation of quantum emitters has already been experimentally demonstrated.
\end{abstract}

\maketitle

\section{Introduction\label{Intro}}
Quantum emitters in semiconductors --- including defect-based color centers, rare-earth dopants, and quantum dots --- have emerged as foundational elements for quantum technologies, offering stable optical transitions, long spin coherence times, and compatibility with scalable photonic architectures. While foundational work has focused on wide-bandgap hosts such as diamond~\cite{aharonovich2016}, recent research has expanded into silicon --- a material combining ultra-low nuclear spin noise, mature CMOS compatibility, and demonstrated spin coherence at both cryogenic and ambient conditions~\cite{saeedi2013,clark2004,zwanenburg2013}. Among the most promising silicon-based emitters are T-centers --- carbon-hydrogen complexes that exhibit telecom-band zero-phonon line (ZPL) emission near 1326 nm and possess spin-½ ground states~\cite{redjem2020}. These centers feature long excited-state lifetimes($\sim$1~\si{\micro\second}) and coherent optical transitions at cryogenic temperatures, making them attractive for integration into silicon photonic circuits~\cite{bergeron2020}.

However, their practical use at room temperature remains constrained by phonon-induced dephasing and non-radiative recombination, which broaden and suppress ZPL emission. To address these limitations, we propose a fundamentally different transduction approach: lifetime-based sensing via modulation of the local density of optical states (LDOS). Unlike conventional schemes based on spectral shifts, this method relies on variations in the emitter’s spontaneous emission rate induced by its electromagnetic environment. Such a detection scheme is inherently more robust to thermal broadening and inhomogeneous line shifts, and remains functional even when the ZPL becomes spectrally unresolved. In other words, even if the emitter rapidly loses coherence (i.e., short $T_2$) and exhibits a broad spectral line, the excited-state population may still decay on a much longer timescale ($T_1$), yielding a measurable lifetime --- the quantity we exploit for sensing. Thermal effects can reduce the quantum efficiency of the emitter, and thus proportionally reduce the achievable sensitivity in our scheme; however, because the detection relies on spontaneous emission lifetime ($T_1$) rather than coherent optical dynamics ($T_2$), the method remains functional even in the presence of strong spectral broadening and dephasing. In this case, the sensitivity scales approximately with the room-temperature quantum efficiency of the emitter, meaning that a moderate reduction in quantum efficiency leads to a proportional, but not prohibitive, reduction in performance. This approach extends beyond T-centers to a broader class of T-like color centers, including group~III --- carbon complexes such as B--C, Al--C, and In--C~\cite{aberl2024,xiong2024}. These centers are predicted to exhibit similar spin and optical properties, with higher structural symmetry (e.g., $C_{2v}$), improved radiative efficiencies, and greater thermal stability --- making them promising candidates for scalable integration with silicon photonics.

While our detection concept does not rely on spin coherence ($T_2$) and is based solely on the excited-state lifetime ($T_1$), demonstrations of exceptionally long $T_2$ in phosphorus donor nuclear spins --- exceeding 39 minutes at room temperature in isotopically purified silicon~\cite{saeedi2013} --- and in $^{29}$Si nuclear spin qubits without isotopic enrichment~\cite{clark2004} affirm silicon’s status as an ultra-low-noise host for quantum emitters. Analogous temperature studies of color centers such as SiV in diamond provide insights into emission linewidths: homogeneous broadening increases with temperature following a $T^3$ dependence, leading to ZPL broadening from $\sim$0.1--0.6~nm at cryogenic temperatures to $\sim$1.3--2~nm at room temperature~\cite{neu2013}. A similar trend is anticipated for silicon-based centers, though experimental verification remains pending. Although room-temperature operation of quantum emitters in silicon remains unproven, no conclusive physical limitation has been established, suggesting potential for future advancement through photonic or material engineering.

In light of these considerations, our proposed sensing scheme --- based on photonic cavity-enhanced lifetime modulation of optically active color centers~\cite{kaupp2016,dovzhenko2020,gritsch2023,kaupp2023,islam2023} --- remains viable under room-temperature conditions provided that the emitter retains a measurable radiative decay channel. While thermal broadening will reduce the maximum attainable Purcell factor, the off-resonance operating regime and lifetime-based detection principle ensure that refractive index sensitivity scales linearly with the retained radiative efficiency. By decoupling sensing performance from spectral resolution constraints, it offers a robust and scalable platform for refractive index detection and single-molecule interaction studies. It is inherently compatible with the future discovery and optimization of silicon-based quantum emitters.

Moreover, this approach occupies a distinctive position among existing quantum sensing methodologies. Unlike nitrogen-vacancy (NV) centers in diamond --- which require intricate spin-state manipulation and suffer from integration challenges --- our method is fully compatible with established silicon photonic fabrication technologies. While fluorescence lifetime modulation via surrounding refractive index modification has previously been explored using NV centers in diamond~\cite{khalid2015}, those approaches typically do not explicitly incorporate the photonic environment via a cavity with defined quality factor $Q$. In contrast, our model analytically links refractive index changes $\Delta n$ to spontaneous emission lifetime through the Purcell factor, which depends on both $\Delta n$ and $Q$. This introduces a new degree of tunability and physical insight, positioning this method as a fundamentally distinct sensing strategy. In addition, while NV-based quantum sensors have demonstrated lifetime-based sensing in magnetometry and microfluidic environments~\cite{sarkar2024,zhang2021,horsthemke2024}, the rigid crystalline nature of diamond hinders their monolithic integration with scalable photonic circuitry. By leveraging T-center and T-like quantum emitters embedded in silicon waveguides, our platform enables site-controlled integration with photonic cavities on a CMOS-compatible substrate. Crucially, it allows the detection of LDOS-modulated lifetime shifts without the need for coherent spin control. This makes the scheme highly attractive for practical, room-temperature sensing and single-molecule interaction detection in compact, scalable devices --- opening a viable path toward quantum-enabled photonic sensing beyond the limitations of spectral resolution and cryogenic operation.

However, it should be noted that, given the generic nature of the proposed approach and the ongoing advances in photonic integrated circuit (PIC) technologies --- such as diamond-on-silicon, hybrid silicon/silicon nitride, and silicon/silicon carbide platforms --- the method can be readily extended to materials in which room-temperature operation of quantum emitters has already been experimentally demonstrated, including diamond, silicon carbide~\cite{lienhard2016}, and silicon nitride~\cite{senichev2021,senichev2022}.  

\begin{figure}[ht]
	\centering
	\includegraphics[width=\linewidth]{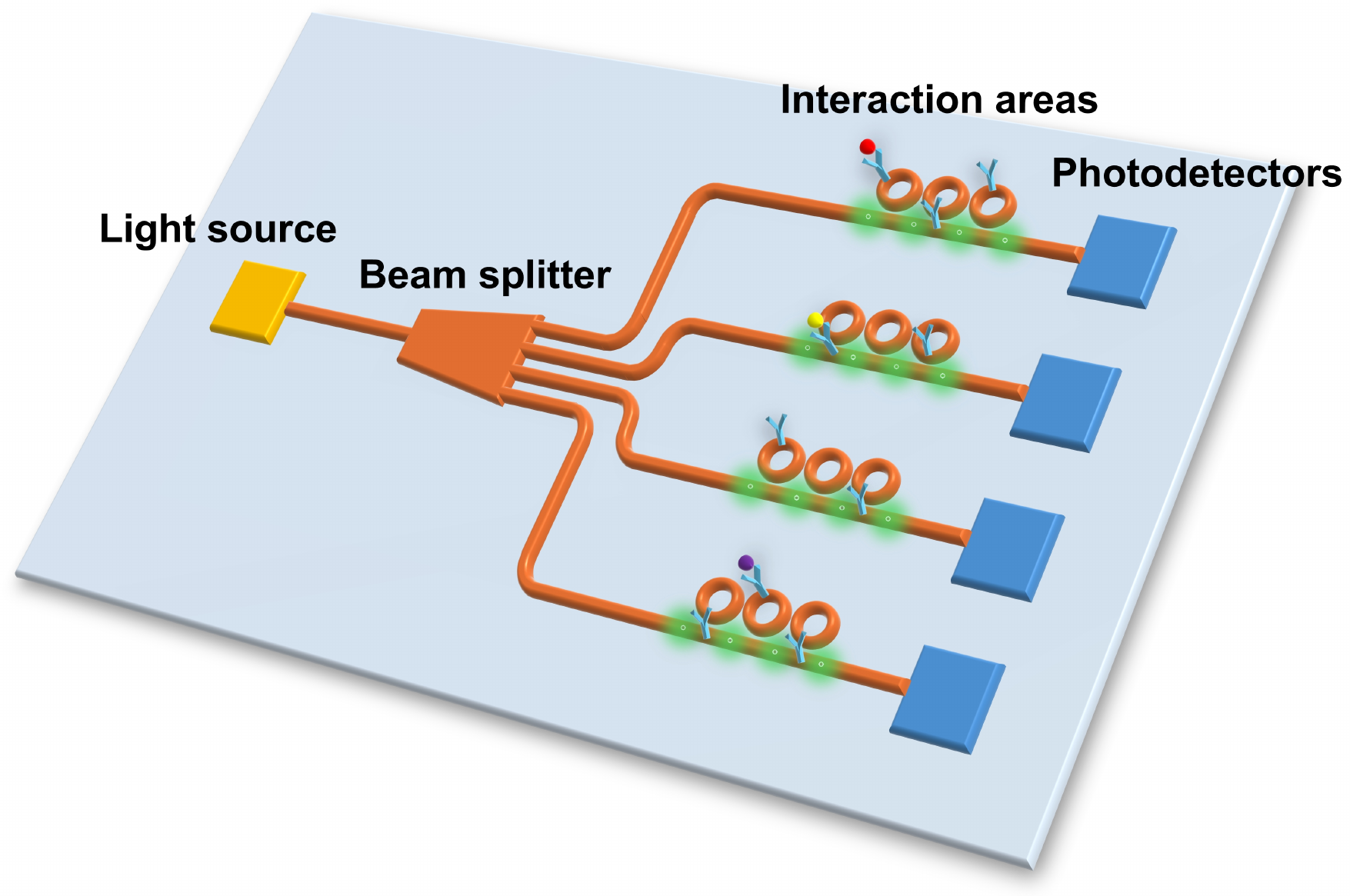}
	\caption{Schematic illustration of a possible multiplexed PIC implementation of the proposed sensing approach, enabling simultaneous detection across multiple channels.}
	\label{fig:figure1}
\end{figure}

Figure~\ref{fig:figure1} illustrates a possible multiplexed realization of the proposed lifetime-based sensing scheme.
Light from a common source is split by an on-chip beam splitter into multiple waveguide channels, each coupled to a high-Q photonic cavity in the “interaction area”. These cavities host quantum emitters functionalized with different biorecognition layers (aptamers, antibodies, etc.), enabling parallel detection of distinct analytes. Analytes can be precisely delivered to the interaction areas by properly designed microfluidic channels. Binding events locally alter the refractive index, modulating the Purcell factor and hence the fluorescence lifetime of the emitters. The emitted photons are collected and directed to dedicated photodetectors for time-correlated single-photon counting, enabling simultaneous, label-free sensing across multiple channels on a single PIC platform. Because the emission originates from cavity-coupled quantum emitters, the platform is fully compatible with efficient photon extraction and does not suffer from the limitations typically associated with high-Q passive cavities.

\section{Theoretical framework and results\label{Theory}}
The spontaneous emission lifetime of a quantum emitter coupled to an optical cavity is governed by the Purcell factor ($F_P$)~\cite{purcell1946}:
\begin{equation} 
	F_P(\vec{r}, \theta) = \frac{3}{4\pi^2} \left( \frac{\lambda}{n} \right)^3 \frac{Q}{V} 
	\left| \frac{\vec{E}(\vec{r}) \cdot \vec{d}}{|\vec{E}_{\text{max}}||\vec{d}|} \right|^2,
	\label{eq:Eq.01}
\end{equation}
where $\lambda$ is the emission wavelength, $n$ is the refractive index of the cavity medium, $Q$ is the quality factor of the cavity, and $V$ is the effective mode volume. 
The term $\vec{E}(\vec{r})$ is the local electric field of the cavity mode at the emitter position $\vec{r}$, $\vec{E}_{\text{max}}$ is the maximum field amplitude in the cavity, and $\vec{d}$ is the unit vector along the emitter's dipole moment.
The angle $\theta$ denotes the angle between the dipole moment $\vec{d}$ and the local electric field vector $\vec{E}(\vec{r})$, such that $\left| \vec{E}(\vec{r}) \cdot \vec{d} \right|^2 = |\vec{E}(\vec{r})|^2 \cos^2 \theta$. This expression accounts for both spatial and orientational mismatch between the emitter and the cavity field, reducing the effective Purcell factor in cases where the emitter is not ideally positioned or aligned.

We start with the ideal case in which the emitter dipole is perfectly aligned with the cavity field, and the emitter is located at the field maximum. In this case, $\theta = 0$ and $\vec{E}(\vec{r}) = \vec{E}_{\text{max}}$, so Eq.~\eqref{eq:Eq.01} simplifies to:
\begin{equation}
	F_P = \frac{3}{4\pi^2} \left( \frac{\lambda}{n} \right)^3 \frac{Q}{V}.
	\label{eq:Eq.02}
\end{equation}
The modified spontaneous emission lifetime $\tau$ is then defined as:
\begin{equation}
	\tau = \frac{\tau_{int}}{F_P},
	\label{eq:Eq.03}
\end{equation}
where $\tau_{int}$ is the intrinsic spontaneous emission lifetime of the emitter in free space or in an unstructured (homogeneous) dielectric. We assume that the optical cavity used to modify the spontaneous emission lifetime of the emitter is optimized to achieve the maximum LDOS and Purcell enhancement. Local changes in the refractive index induce a spectral detuning of the cavity resonance. This detuning reduces the LDOS at the emitter's location, thereby suppressing the spontaneous emission rate and resulting in an increase in the emitter's lifetime. This lifetime shift can be detected using time-correlated single-photon counting (TCSPC), forming the basis for a label-free sensing modality. Given this, we estimate the effect of a refractive index change on the spontaneous emission lifetime of a quantum emitter coupled to a photonic cavity. A local change in the refractive index alters the effective index of the cavity mode, $n_\mathrm{eff}$, leading to a shift in cavity’s resonance angular frequency $\omega_\mathrm{cav}$. In the regime of small perturbations, the shift in the resonance frequency can be expressed using first-order perturbation theory as:
\begin{equation}
	\frac{\Delta \omega}{\omega_\mathrm{cav}} \approx -\frac{1}{2} \frac{\int \Delta \varepsilon(\vec{r}) \left| \vec{E}(\vec{r}) \right|^2 dV}{\int \varepsilon(\vec{r}) \left| \vec{E}(\vec{r}) \right|^2 dV},
	\label{eq:Eq.04}
\end{equation}
where $\Delta \varepsilon(\vec{r}) = 2 n(\vec{r}) \Delta n(\vec{r})$ is the permittivity perturbation induced by the local refractive index change $\Delta n(\vec{r})$, and $\vec{E}(\vec{r})$ is the cavity mode electric field. It should be noted here that in this work, we focus on refractive index variations arising from molecular binding events, while other possible $\Delta n$ sources (e.g., temperature fluctuations) are considered parasitic and should be minimized in practical implementations. By assuming $\Delta \varepsilon = 2n \Delta n \approx 2n_{\mathrm{eff}} \Delta n$, we can then simplify Eq.~\eqref{eq:Eq.04} to:
\begin{equation}
	\Delta \omega = \omega_{\mathrm{cav}} \frac{\Delta n}{n_{\mathrm{eff}}}.
	\label{eq:Eq.05}
\end{equation}
Depending on cavity type, this detuning lowers the spectral overlap between the emitter (e.g., T-center) and the cavity mode, reducing the Purcell factor. The Purcell factor $F_P$ under detuning follows a Lorentzian dependence:
\begin{equation}
	F_P(\Delta \omega) = \frac{F_{P,\mathrm{max}}}{1 + \left( \frac{2Q \Delta \omega}{\omega_\mathrm{cav}} \right)^2},
	\label{eq:Eq.06}
\end{equation}
where $\Delta \omega = \omega_\mathrm{emitter} - \omega_\mathrm{cav}$ is the detuning. If the cavity is on-resonance in the initial state: $\Delta \omega = 0$, and $F_P = F_{P,\mathrm{max}}$, and lifetime is at its minimum ($\tau = \tau_{int} / F_P$). Then, a small change in $F_P$ leads to the following relative lifetime change of the emitter:
\begin{equation}
	\frac{\Delta \tau}{\tau} \approx \left( \frac{2Q \Delta n}{n_\mathrm{eff}} \right)^2.
	\label{eq:Eq.07}
\end{equation}

\begin{figure*}[ht]
	\centering
	\includegraphics[width=\linewidth]{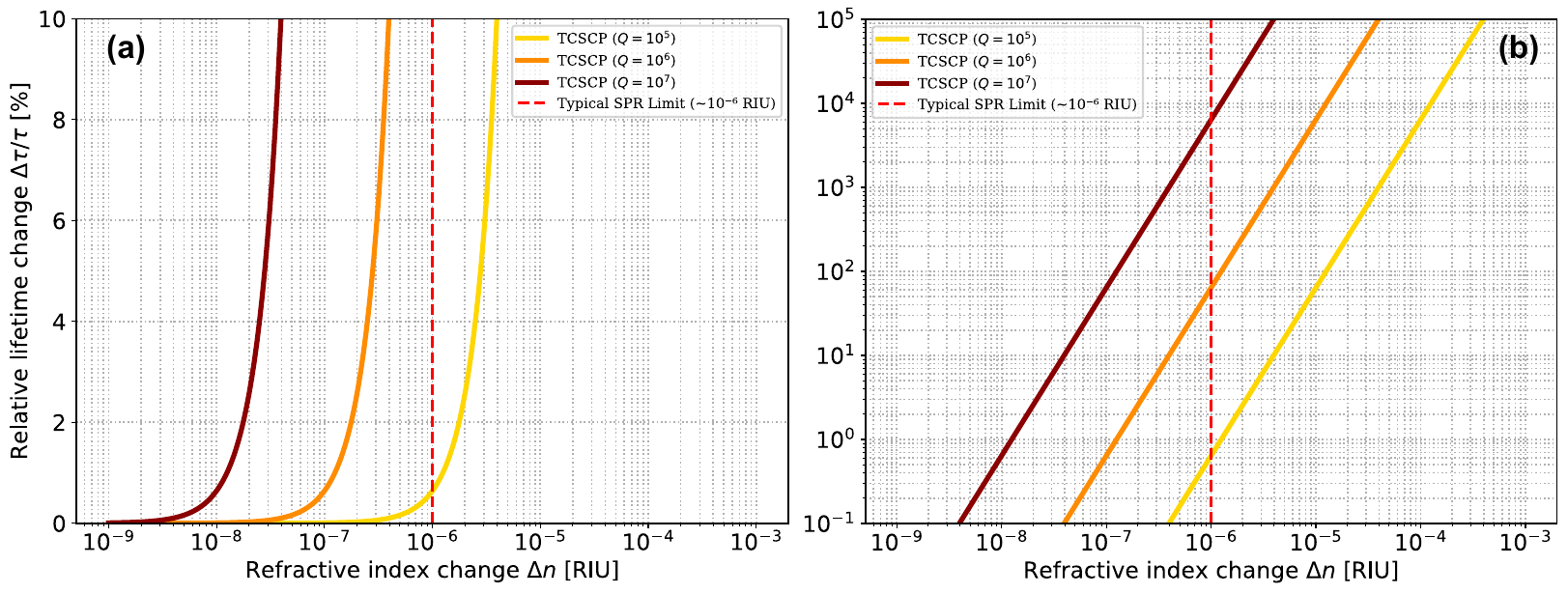}
	\caption{Simulated lifetime change ($\Delta \tau / \tau$) versus refractive index change ($\Delta n$) (i.e., sensitivity): (a) linear scale for $\Delta \tau / \tau$ from 0 to 10\% and logarithmic scale for $\Delta n$; (b) logarithmic scale for $\Delta \tau / \tau$ from $10^{-1}$ to $10^{5}$ and logarithmic scale for $\Delta n$.}
	\label{fig:figure2}
\end{figure*}

For parametric evaluation of Eq.~\eqref{eq:Eq.07}, we assume the following values: $Q = 10^5$, $10^6$, or $10^7$ and $n_\mathrm{eff} = 2.5$. Simulated relative lifetime change $\Delta \tau / \tau$ versus refractive index change $\Delta n$ comparing the TCSPC approach with the state-of-the-art plasmonic detection limit is shown in Figure~\ref{fig:figure2}, where Figure~\ref{fig:figure2}a features linear scale for the $\Delta \tau / \tau$ ranging from 0 to 10\% and logarithmic scale for the $\Delta n$, while Figure~\ref{fig:figure2}b --- both in logarithmic scale.

According to Figure~\ref{fig:figure2}, in cases of the cavity with $Q = 10^5$ in the on-resonance state, achieving the state-of-the-art sensitivity of the plasmonic measurements ($10^{-6}$~RIU) requires the ability of measuring the relative lifetime changes of less than 1\%. Solving Eq.~\eqref{eq:Eq.07} for $\Delta n$ and substituting these values shows that we are potentially sensitive to refractive index changes as small as $\Delta n \approx 4.68 \times 10^{-7}$ ($Q = 10^5$), which may compete with or even surpass the sensitivity of state-of-the-art plasmonic devices. In general, for a photonic cavity with $Q = 10^5$, achieving a relative lifetime change $\Delta \tau / \tau$ below 0.64\% (intersection of the yellow solid curve and red dashed line in Figure~\ref{fig:figure2}) would enable the proposed system to surpass the $10^{-6}$~RIU sensitivity limit typical of plasmonic sensors.

\begin{figure*}[ht]
	\centering
	\includegraphics[width=\linewidth]{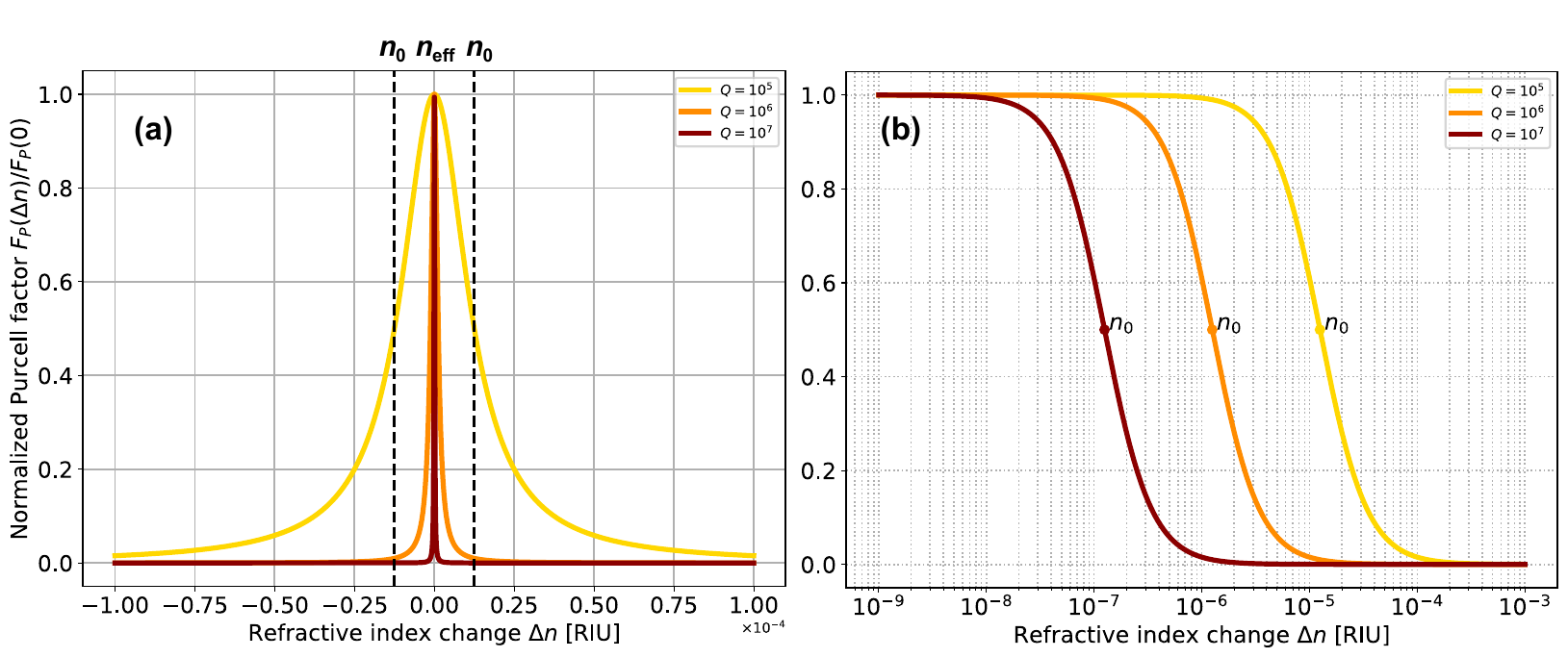}
	\caption{Simulated normalized Purcell factor versus refractive index change $\Delta n$: (a) linear scale for $\Delta n$ from $-10^{-4}$ to $+10^{-4}$; (b) log scale for $\Delta n$ from $10^{-9}$ to $10^{-3}$.}
	\label{fig:figure3}
\end{figure*}

However, at resonance ($\Delta \omega = 0$), the Purcell factor is at its maximum, but its derivative with respect to detuning is zero. Therefore, in the immediate vicinity of $\Delta n = 0$ (i.e., near the point $n_\mathrm{eff}$, see Figures~\ref{fig:figure2}a and~\ref{fig:figure3}), the dependence of $\Delta \tau / \tau$ on $\Delta n$ becomes non-linear (specifically, quadratic; see Eq.~\eqref{eq:Eq.07}), making it unsuitable for applications that require quantitative detection of small refractive index changes.
A more effective approach is to shift the emitter-cavity system off-resonance, to the point where the slope of the Lorentzian Purcell factor is maximized --- i.e., at detuning by $\Delta \omega = \pm \kappa / 2 = \pm \omega_\mathrm{cav} / (2Q)$ where $\kappa$ is the cavity decay rate. This $\Delta \omega$ detuning relates to the following required shift in the refractive index from the on-resonance point, $n_\mathrm{shift} = n_0 - n_\mathrm{eff} = \pm n_\mathrm{eff} / 2Q$, where $n_0$ is the off-resonance point we want to work around: $n_0 = n_\mathrm{eff} \pm n_\mathrm{eff} / 2Q = n_\mathrm{eff} (1 \pm 1 / 2Q)$. Here, the relative lifetime change becomes linearly dependent on $\Delta n$ for small perturbations, enabling maximum sensitivity and restoring a wide, usable dynamic range (see Figure~\ref{fig:figure3}). At these points (marked by vertical black lines in Figure~\ref{fig:figure3}a for $Q = 10^5$ and by color dots in Figure~\ref{fig:figure3}b for all curves), the lifetime of quantum emitters is not at its minimum, but its sensitivity to detuning is maximal. This trade-off introduces a moderate ($\sim$50\%) reduction in Purcell enhancement but enables a linear, tunable sensing regime suitable for detecting small $\Delta n$ changes. This strategy also introduces a modest reduction in intensity (since emission rate is lower), but dramatically improves the usable dynamic range. The reduction in intensity can be mitigated by better photodetectors, longer integration, and higher emitter density. Especially for long-lived emitters like T-centers, this intensity loss is well tolerated, and the resulting linear response regime is highly advantageous for sensing applications.

We next derive the expression for the detection limit $\Delta n_\mathrm{opt}$ related to these optimal detection points for each $Q$ value. The emission lifetime $\tau$ at the shifted point $n_0$ can be written as:
\begin{equation}
	\tau(n_0) = \frac{\tau_\mathrm{eff}}{1 + \left( 2Q \frac{(n_0 - n_\mathrm{eff})}{n_\mathrm{eff}} \right)^2},
	\label{eq:Eq.08}
\end{equation}
where $\tau_\mathrm{eff}$ is the lifetime of the quantum emitter in the on-resonance state. In the approximation of $\Delta n \ll n_\mathrm{shift}$, the emission lifetime change around $n_0$ can be expressed as follows:
\begin{align}
	\frac{d\tau(n_0)}{dn_0} &= -\frac{\tau_\mathrm{eff}}{\left[ 1 + \left( 2Q \frac{(n_0 - n_\mathrm{eff})}{n_\mathrm{eff}} \right)^2 \right]^2} \cdot 2 \cdot 2Q \frac{(n_0 - n_\mathrm{eff})}{n_\mathrm{eff}} \cdot \frac{2Q}{n_\mathrm{eff}} \notag \\
	&= -4\tau_0 \frac{Q^2}{n_\mathrm{eff}^2} n_\mathrm{shift} \notag \\
	&= \pm 2\tau_0 \frac{Q}{n_\mathrm{eff}} = \pm 4\tau_\mathrm{eff} \frac{Q}{n_\mathrm{eff}},
	\label{eq:Eq.09}
\end{align}
here $\tau_0 = \tau(n_0) = 2\tau(n_\mathrm{eff}) = 2\tau_\mathrm{eff}$. In the vicinity of $n_0$, we can write:
\begin{align}
	\tau(n_0 + \Delta n) &= \tau(n_0) + \frac{d\tau(n_0)}{dn_0} \Delta n \notag \\
	&= \tau_0 \pm 2\tau_0 \frac{Q}{n_\mathrm{eff}} \Delta n \notag \\
	&= \tau_0 \left( 1 \pm 2 \frac{Q}{n_\mathrm{eff}} \Delta n \right).
	\label{eq:Eq.10}
\end{align}
From Eq.~\eqref{eq:Eq.10} it follows that:
\begin{equation}
	\Delta \tau = \pm 2\tau_0 \frac{Q}{n_\mathrm{eff}} \Delta n,
	\label{eq:Eq.11}
\end{equation}
and consequently:
\begin{equation}
	\left| \frac{\Delta \tau / \tau_0}{\Delta n} \right| = 2 \cdot \frac{Q}{n_\mathrm{eff}}.
	\label{eq:Eq.12}
\end{equation}

Eq.~\eqref{eq:Eq.12} gives the slope of the Lorentzian response curve at the point of maximum sensitivity $n_0$. According to Eq.~\eqref{eq:Eq.12}, for a typical effective refractive index of $n_\mathrm{eff} = 2.5$, the sensitivity slopes are depicted in Table~\ref{tab:Tab1}.

\begin{table}[ht]
	\centering
	\caption{Sensitivity slopes as a function of the $Q$ factor of the cavity.}
	\label{tab:Tab1}
	\renewcommand{\arraystretch}{1.2}
	\begin{tabular}{c@{\hspace{0.5cm}}c}
		\hline
		\textbf{Q} & \textbf{$|\Delta\tau/\tau|$ per $\Delta n$ (RIU$^{-1}$)} \\
		\hline
		$10^5$ & $8 \times 10^4$ \\
		$10^6$ & $8 \times 10^5$ \\
		$10^7$ & $8 \times 10^6$ \\
		\hline
	\end{tabular}
\end{table}
These values represent the slope of the relative lifetime change per unit refractive index change, measured at the point of optimal detuning ($\Delta\omega = \kappa/2$). This slope enables conversion between $\Delta n$ and the required measurable $\Delta \tau / \tau_0$ for a given $Q$. Eq.~\eqref{eq:Eq.12} can be re-written for $\Delta n_{\text{opt}}$ as:
\begin{equation}
	\Delta n_{\text{opt}} = \frac{\left( \Delta \tau / \tau_0 \right)}{2Q} \cdot n_{\text{eff}}.
	\label{eq:Eq.13}
\end{equation}
Eq.~\eqref{eq:Eq.13} estimates the minimum detectable $\Delta n_{\text{opt}}$ for a given temporal resolution. According to Eq.~\eqref{eq:Eq.13}, the detection limit of the proposed refractive index sensing approach in the off-resonance regime, $\Delta n_{\text{opt}}$, is limited by the $Q$ value of the photonic cavity and the resolution capability $\Delta \tau / \tau_0$ of the TCSPC system used. For instance, if a TCSPC system can resolve a 3\% change in lifetime ($\Delta \tau / \tau_0 = 0.03$), then the minimum detectable $\Delta n_{\text{opt}}$ would be approximately $3.75 \times 10^{-7}$~RIU for $Q = 10^5$, $3.75 \times 10^{-8}$~RIU for $Q = 10^6$, and $3.75 \times 10^{-9}$~RIU for $Q = 10^7$, assuming $n_{\text{eff}} = 2.5$. In addition, we calculated the linear dynamic range (LDR) for each $Q$ value as the span of refractive index changes $\Delta n$ over which the relative lifetime change $\Delta \tau / \tau_0$ remains approximately linear (with a deviation less than $\pm$10\% from linear behavior) around the $n_0$ point. The LDR calculation details can be found in the Annex (see Supporting Information). The Python script and accompanying description for LDR calculation are available for download (see the Data Availability section at the end of the paper). These data are summarized in Table~\ref{tab:Tab2}.
\begin{table}[ht]
	\centering
	\caption{Minimum detectable $\Delta n_{\text{opt}}$ and LDR width as a function of $Q$.}
	\label{tab:Tab2}
	\renewcommand{\arraystretch}{1.5}
	\begin{tabular}{c@{\hspace{0.5cm}}c@{\hspace{0.5cm}}c}
		\hline
		\textbf{Q} & \textbf{Minimum $\Delta n_{\text{opt}}$ [RIU]} & \textbf{LDR width [RIU]} \\
		\hline
		$10^5$ & $3.75 \times 10^{-7}$ & $1.12 \times 10^{-5}$ \\
		$10^6$ & $3.75 \times 10^{-8}$ & $1.12 \times 10^{-6}$ \\
		$10^7$ & $3.75 \times 10^{-9}$ & $1.12 \times 10^{-7}$ \\
		\hline
	\end{tabular}
\end{table}

At this point we must address the consequences of the idealizations made when transitioning from Eq.~\eqref{eq:Eq.01} to Eq.~\eqref{eq:Eq.02}. In practice, achieving perfect spatial alignment and dipole orientation of quantum emitters in silicon photonic cavities is highly challenging with current technology. For example, standard techniques such as ion implantation followed by thermal annealing~\cite{macquarrie2021} result in emitters that are distributed in space with a resolution no better than 20 nm and possess stochastic dipole orientations relative to the cavity mode field. At the same time, the Purcell factor $F_P$ is strongly dependent not only on the emitter's spectral resonance with the cavity mode but also on its position within the cavity and the relative orientation of its dipole moment with respect to the local electric field~\cite{zalogina2017}. As a result, Eq.~\eqref{eq:Eq.12}, which describes the slope of the Lorentzian lifetime response curve at the point of maximum sensitivity $n_0$, reflects an idealized scenario assuming maximum Purcell coupling --- i.e., the emitter is located at the field antinode and its dipole is perfectly aligned with the local field vector. In realistic implementations, however, spatial and orientational averaging must be taken into account.

To capture this deviation, we introduce an effective coupling efficiency factor $\eta_\mathrm{eff} \in [0,1]$, which quantifies the reduction of the average Purcell factor due to positional and angular mismatch. This factor can be expressed as:
\begin{equation}
	\eta_\mathrm{eff} = \left\langle \cos^2 \theta \right\rangle \cdot \frac{\left\langle \left| \vec{E}(\vec{r}) \right|^2 \right\rangle}{\left| \vec{E}_{\mathrm{max}} \right|^2},
	\label{eq:Eq.14}
\end{equation}
where \(\left\langle \cos^2 \theta \right\rangle = \int_0^{\pi} \cos^2 \theta \sin \theta \, d\theta \, / \, \int_0^{\pi} \sin \theta \, d\theta = \frac{1}{3}\) represents the average projection of the dipole on the field direction for isotropic orientation in three dimensions, and $\left\langle |\vec{E}(\vec{r})|^2 \right\rangle / |\vec{E}_{\mathrm{max}}|^2$ accounts for the average field intensity experienced by the emitters relative to the maximum. If the cavity mode field has the form \( E(x) \sim \cos(kx) \), then the intensity is given by \( |E(x)|^2 \sim \cos^2(kx) = \left(1 + \cos(2kx)\right)/2 \) (here $k = \dfrac{2\pi}{\lambda}$ is the wavevector of the cavity mode and $\lambda$ is the wavelength of the mode). When averaging over the spatial coordinate \( x \), we obtain \( \langle \cos^2(kx) \rangle = 1/2 \). According to Eq.~\eqref{eq:Eq.14}, this yields $\eta_\mathrm{eff} \sim 1/6$ as a reasonable estimate. Taking this factor into account, the realistic slope of the Lorentzian lifetime response curve becomes:
\begin{equation}
	\left| \frac{\Delta \tau / \tau_0}{\Delta n} \right|_\mathrm{realistic} = \eta_\mathrm{eff} \cdot \left( \frac{2Q}{n_\mathrm{eff}} \right) \approx \frac{1}{6} \cdot \left( \frac{2Q}{n_\mathrm{eff}} \right),
	\label{eq:Eq.15}
\end{equation}
which can be interpreted as the effective sensitivity of the system in realistic experimental conditions. While this reduction is not negligible, it does not negate the viability of the sensing scheme. On the contrary, it highlights the importance of modeling the average response and sets a benchmark for achievable performance using ensembles of implanted quantum emitters. Furthermore, the inherent sensitivity of long-lived emitters (such as T-centers) combined with the robustness of TCSPC techniques ensures that even a fraction of the ideal Purcell response can yield detectable and quantifiable lifetime shifts for practical sensing applications. Notably, this reduction in effective sensitivity due to spatial and orientational averaging can be partially compensated by employing cavities with one order higher $Q$, thereby restoring comparable performance.

It is worth noting that while Eq.~\eqref{eq:Eq.14} provides a useful approximation for ensemble coupling efficiency under isotropic and homogeneous assumptions, an alternative form involving an explicit sum over emitter positions and orientations can be written as:
\begin{equation}
	\eta_{\text{eff}} = \frac{1}{N} \sum_{i=1}^N \cos^2\theta_i \cdot \frac{|\vec{E}(\vec{r}_i)|^2}{|\vec{E}_{\text{max}}|^2},
	\label{eq:Eq.16}
\end{equation}
where $\theta_i$ is the angle between the $i$-th emitter's dipole moment and the local electric field, and $\vec{r}_i$ is its spatial position. This formulation reduces to Eq.~\eqref{eq:Eq.14} when spatial and angular averaging is applied under the assumption of uniform emitter distribution and isotropic orientation. Such an emitter-resolved form can be particularly useful in the context of precise emitter engineering --- for example, when emitters are deterministically placed at specific field maxima or aligned to maximize coupling. In the limit of a continuous distribution $\rho(\vec{r})$, the average coupling efficiency may be expressed as:
\begin{equation}
	\eta_{\text{eff}} = \frac{1}{N} \int_V \rho(\vec{r}) \left\langle \cos^2 \theta \right\rangle \cdot \frac{|\vec{E}(\vec{r})|^2}{|\vec{E}_{\text{max}}|^2} \, dV,
	\label{eq:Eq.17}
\end{equation}
where $N = \int_V \rho(\vec{r}) \, dV$ is the total number of emitters. This continuous formalism remains normalized between 0 and 1, and allows for accurate modeling of spatially varying emitter densities and orientation distributions, enabling more precise predictions in engineered photonic environments.

\begin{figure}[ht]
	\centering
	\includegraphics[width=\linewidth]{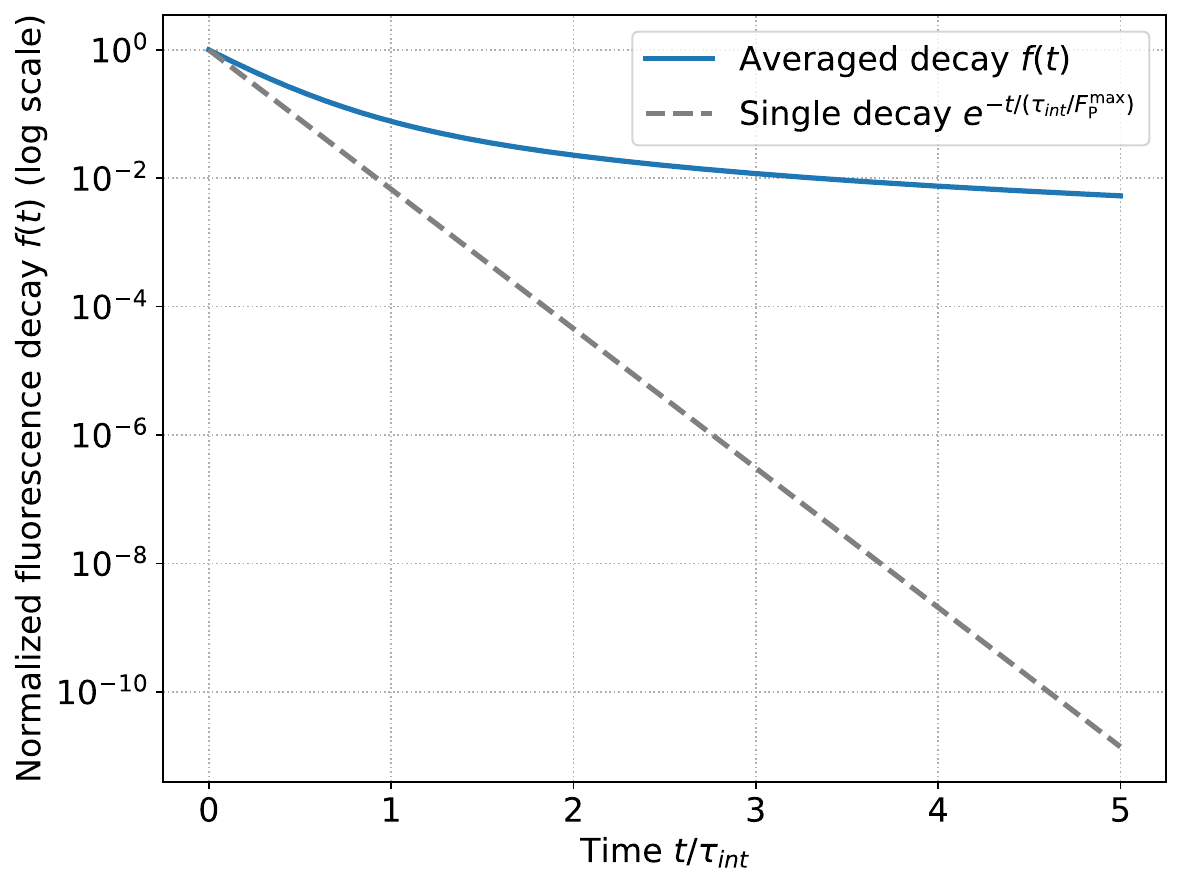}
	\caption{Normalized fluorescence decay $f(t)$ (log scale) versus time $t / \tau_{int}$: $F_P^{\text{max}}=5$, $\tau_{int}=1\mu$s.}
	\label{fig:figure4}
\end{figure}

From a practical point of view, however --- e.g., for measuring fluorescence emission --- the following aspects are more important. The fluorescence intensity decays exponentially, and the decay rate differs for different emitters depending on their position and orientation. Therefore, it is essential to determine the fluorescence decay averaged over both the orientation and position of an ensemble of the emitters. The normalized fluorescence decay function $f(t)$, accounting for orientation and spatial averaging, can be expressed as:
\begin{equation}
	f(t) = \frac{ \displaystyle \int_0^{\pi} \sin\theta \, d\theta \int_0^{\lambda/2} \cos^2(kx) \, \exp\left(-\frac{t}{\tau(\theta, x)}\right) dx }{ \displaystyle \int_0^{\pi} \sin\theta \, d\theta \int_0^{\lambda/2} dx },
	\label{eq:Eq.18}
\end{equation}
where the position- and orientation-dependent lifetime $\tau(\theta, x)$ is defined as:

\begin{equation}
	\tau(\theta, x) = \frac{\tau_\text{int}}{F_P^{\text{max}} \cos^2(kx) \cos^2(\theta)}.
	\label{eq:Eq.19}
\end{equation}
Eq.~\eqref{eq:Eq.18} models the normalized fluorescence decay curve $f(t)$ for an ensemble of quantum emitters embedded in a cavity. Each emitter may be located at a different position within the cavity and oriented at a different angle with respect to the local electric field. These factors modulate the spontaneous emission rate due to the Purcell effect. The model accounts for this by averaging the decay function over both spatial and orientational distributions. Eq.~\eqref{eq:Eq.18} was solved numerically, and Figure~\ref{fig:figure4} shows dependence of the normalized fluorescence decay $f(t)$ on the time $t / \tau_{int}$. 

According to Figure~\ref{fig:figure4}, the deviation of the averaged fluorescence decay curve from a straight line in logarithmic scale confirms its non-exponential nature. This behavior reflects the distribution of local lifetimes across the ensemble due to variations in emitter position and dipole orientation, resulting in a multi-exponential decay profile. Such effects are well known in complex photonic and biological systems~\cite{becker2023}, where spatial and structural inhomogeneities introduce deviation from a single-exponential decay and lead to a broadening of the decay dynamics.

While this may seem trivial --- since even the sum of just two exponentials already deviates from a single-exponential function --- it has practical implications. In particular, extracting lifetime information or classifying decay behavior becomes nontrivial when the signal contains a continuum of lifetime contributions. In this context, machine learning and AI-based tools are promising. They can be trained to identify patterns in TCSPC data, rapidly approximate multi-lifetime combinations, or directly classify decay signals without rigid model assumptions.

Furthermore, two practical aspects are important when interpreting experimental decay curves. First, while fluorescence signals are measured in absolute units (e.g., photon counts per time bin), the relative uncertainty increases as the signal decreases. This means that in the "tail" of the decay, even small fluctuations in absolute counts can result in significant relative error. Second, as the decay slows at longer times --- clearly visible in the logarithmic scale --- larger time bin widths may be used (i.e., coarser binning), which helps reduce data volume. However, achieving sufficient signal-to-noise ratio in this region still requires extended acquisition time or higher photon collection efficiency.

To summarize, model provided by the Eq.~\eqref{eq:Eq.18} realistically predicts a non-exponential decay curve due to variations in emitter positions and orientations. It provides a physically accurate description of ensemble fluorescence in cavity-enhanced settings and is directly applicable to the interpretation of TCSPC experiments. It can also support fitting algorithms and machine learning-based lifetime analysis in nanophotonic biosensors.

From these considerations, it is clear that the proposed sensing mechanism can be considered as future-forward, targeting best-in-class performance, as it can potentially resolve $\Delta n$ changes down to $10^{-9}$~RIU. Even lower values can be readily achievable for cavities with $Q$-factors $> 10^7$, achievable via state-of-the-art silicon photonics cavity designs~\cite{quan2011}. This corresponds to a minimum detectable refractive index shift of $< 10^{-9}$~RIU, outperforming plasmonic counterparts and unlocking single-molecule sensitivity in compact, integrated devices.

Another practical advantage of operating the sensor off-resonance is the significant relaxation of the technical requirements on the time-resolved detection system. At exact resonance ($\Delta\omega = 0$), the relative lifetime change in response to small refractive index shifts can be extremely low (see Figure~\ref{fig:figure2}a) --- often below 1\% --- which necessitates ultra-high-resolution TCSPC instrumentation. In contrast, at the optimal detuning point ($\Delta\omega = \kappa/2$), the slope of the Purcell factor with respect to detuning is maximized, enabling much larger and more easily measurable lifetime changes for the same $\Delta n$. As a result, the proposed off-resonance sensing strategy enables high sensitivity while maintaining a practical detection limit for the TCSPC system.

While higher-$Q$ cavities generally enhance sensitivity, they do not always translate into stricter requirements for the TCSPC system—especially when operating off-resonance. At higher $Q$ values, the Purcell effect becomes stronger, leading to shorter emitter lifetimes in the initial state. However, if the system is intentionally operated at the point of maximum slope of the Purcell curve (i.e., off-resonance at $\Delta\omega = \kappa/2$), the relative lifetime changes become significantly larger and more linear, even for small refractive index shifts. This approach reduces the need for ultra-high temporal resolution, as detecting a few percent change in lifetime (instead of sub-percent) is well within the capability of modern TCSPC systems. At the same time, increasing $Q$ introduces other challenges, such as tighter fabrication tolerances, greater thermal sensitivity, and more demanding integration procedures. The choice of $Q$ should therefore balance sensitivity, fabrication feasibility, and instrumentation requirements.

\section{Advantage of long-lived emitters like T-centers\label{Effective lifetime}}
\begin{figure}[ht]
	\centering
	\includegraphics[width=\linewidth]{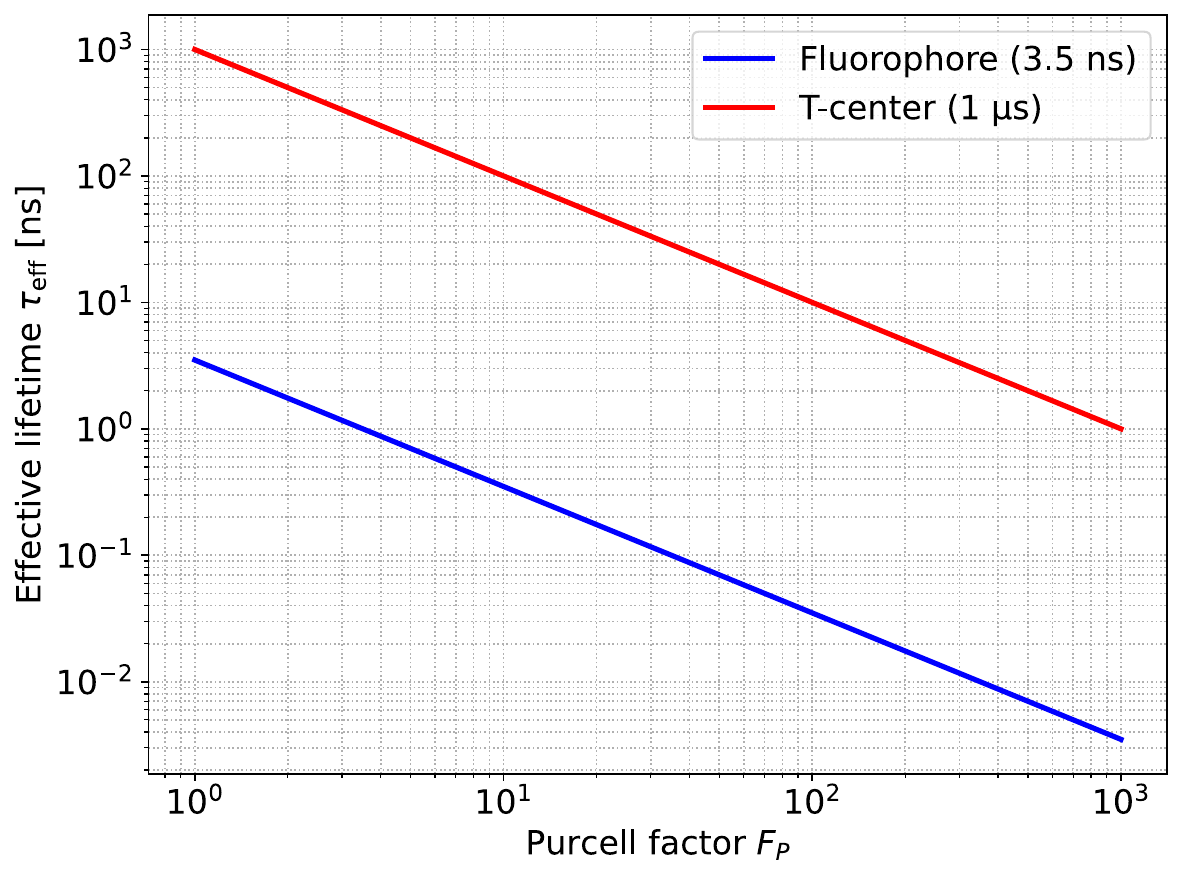}
	\caption{Comparison of effective lifetimes $\tau_{\mathrm{eff}}$ of a typical fluorophore ($\sim$3.5~ns intrinsic) and a T-center with an uncoupled lifetime of 1~$\mu$s, as a function of Purcell factor.}
	\label{fig:figure5}
\end{figure}

One of the key technical challenges in high-$Q$ cavity-enhanced sensing is the requirement for very high time resolution in TCSPC systems, especially when using fast-emitting fluorophores like organic dyes ($\tau \sim 3.5$~ns). In those cases, small relative lifetime changes (e.g., 0.1--0.3\%) correspond to absolute shifts in the range of 3--10~ps --- demanding the use of specialized ultrafast detection equipment. However, T-centers in silicon offer a unique advantage due to their long natural lifetimes, often in the microsecond range. For instance, a T-center with a cavity-unaffected (uncoupled) lifetime of 1~$\mu$s~\cite{islam2023}, when coupled to a photonic cavity with a moderate Purcell factor (e.g., $F_P = 5$), will exhibit a reduced lifetime of about 200~ns. In this regime, even a 0.1\% lifetime change corresponds to a shift of 200~ps --- easily detectable with standard TCSPC systems. This greatly relaxes the instrumentation requirements and enables detection of ultra-small refractive index changes, making T-centers a highly attractive emitter platform for integrated biosensing using cavity-enhanced lifetime detection. Figure~\ref{fig:figure5} shows a comparison of effective lifetimes $\tau_{\mathrm{eff}}$ of a typical fluorophore (3.5~ns intrinsic) and a T-center with an uncoupled lifetime of 1~$\mu$s, as a function of Purcell factor. T-centers maintain a large $\tau_{\mathrm{eff}}$ ($\sim$1~ns) even under high Purcell enhancement ($F_P = 10^3$), easing TCSPC requirements.

In the context of off-resonance sensing, where the system is operated at the point of maximum slope in the Purcell response curve, a relative change in lifetime of a few percent (e.g., 3\%) is both realistic and desirable. This linear regime simplifies signal interpretation and reduces the burden on instrumentation. The absolute lifetime change required for detection is approximately \(\Delta \tau = 0.03 \cdot \tau_{\mathrm{eff}}\), where \(\tau_{\mathrm{eff}}\) is the cavity-enhanced lifetime. This requirement depends on the emitter type and the cavity-enhanced lifetime $\tau_{\mathrm{eff}}$, which is reduced from the intrinsic value via the Purcell effect as shown in Table~\ref{tab:Tab3}.
\begin{table*}[ht]
	\centering
	\caption{Estimated lifetime and $\Delta \tau$ requirements for fluorescent dyes (3.5~ns) and T-centers (1~$\mu$s) under various $Q$ values assuming off-resonance sensing.}
	\label{tab:Tab3}
	\renewcommand{\arraystretch}{1.5}
	\begin{tabular}{c@{\hspace{0.5cm}}c@{\hspace{0.5cm}}c}
		\hline
		\textbf{Q} & \textbf{Fluorescent Dyes} & \textbf{T-Centers} \\
		\hline
		$10^5$ & 
		\makecell[l]{Est. $F_P$: $\sim$10 \\
			$\tau_{\mathrm{eff}}$: $\sim$0.35~ns \\
			$\Delta \tau$ (3\%): $\sim$10.5~ps \\
			Required TCSPC: $\sim$10–20~ps (high-end)} & 
		\makecell[l]{Est. $F_P$: $\sim$10 \\
			$\tau_{\mathrm{eff}}$: $\sim$100~ns \\
			$\Delta \tau$ (3\%): $\sim$3~ns \\
			Required TCSPC: $\sim$2–5~ns (standard)} \\
		\hline
		$10^6$ & 
		\makecell[l]{Est. $F_P$: $\sim$50 \\
			$\tau_{\mathrm{eff}}$: $\sim$0.07~ns \\
			$\Delta \tau$ (3\%): $\sim$2.1~ps \\
			Required TCSPC: $<$10~ps (cutting-edge)} & 
		\makecell[l]{Est. $F_P$: $\sim$50 \\
			$\tau_{\mathrm{eff}}$: $\sim$20~ns \\
			$\Delta \tau$ (3\%): $\sim$600~ps \\
			Required TCSPC: $\sim$500~ps–1~ns (standard)} \\
		\hline
		$10^7$ & 
		\makecell[l]{Est. $F_P$: $\sim$100 \\
			$\tau_{\mathrm{eff}}$: $\sim$0.035~ns \\
			$\Delta \tau$ (3\%): $\sim$1.0~ps \\
			Required TCSPC: $<$5~ps (advanced only)} & 
		\makecell[l]{Est. $F_P$: $\sim$100 \\
			$\tau_{\mathrm{eff}}$: $\sim$10~ns \\
			$\Delta \tau$ (3\%): $\sim$300~ps \\
			Required TCSPC: $\sim$200–500~ps (standard)} \\
		\hline
	\end{tabular}
\end{table*}
As shown in Table~\ref{tab:Tab3}, fluorescent dyes require very high timing resolution (20~ps or better) for high-$Q$ cavities, making them challenging to use without ultrafast detection systems. In contrast, T-centers, even when strongly coupled to high-$Q$ cavities, maintain effective lifetimes in the tens to hundreds of nanoseconds. This translates into much larger absolute lifetime changes, which can be resolved using standard TCSPC systems with time resolution in the hundreds of picoseconds to a few nanoseconds. This distinction makes T-centers a highly attractive choice for scalable and practical high-sensitivity sensing. It should be noted that the “Required TCSPC” values in Table III are provided as illustrative benchmarks of commercially available temporal resolution and should not be interpreted as strict physical limits. In practice, the detectability of small fractional lifetime changes is determined primarily by photon statistics and signal-to-noise ratio. Coarser binning may be employed to improve photon counts per bin without significantly degrading the ability to resolve changes in decay slope.

\section{Comparison with other approaches based on resonance wavelength shifts\label{Comparison}}
In addition to comparing our approach with plasmonic measurements, it is also important to evaluate it against other sensing methods based on resonance wavelength shifts. As a benchmark, we consider the paper by Armani et al.~\cite{armani2007}, in which an ultrahigh quality factor ($Q > 10^8$) whispering-gallery microcavity (WGM) was used for direct, label-free, single-molecule detection of interleukin-2 in serum. The WGM cavity was fabricated in a thermally oxidized SiO$_2$ layer on top of a silicon wafer. The silica surface was functionalized to specifically bind the target molecule, with detection based on shifts in the resonant wavelength. Single-molecule sensitivity was confirmed by observing discrete binding events that induced measurable changes in the resonance frequency.

We now estimate the minimum detectable $\Delta n$ value that was observed in the paper and compare it with the expected values achievable using our proposed approach. In the benchmark paper, discrete resonance shifts on the order of $\sim$50 femtometers (fm) per single-molecule binding event were observed. The relationship between the resonance wavelength shift $\Delta \lambda$ and the change in the effective refractive index $\Delta n$ can be estimated as:
\begin{equation}
	\frac{\Delta \lambda}{\lambda_r} \approx \frac{\Delta n}{n_{\text{eff}}}.
	\label{eq:Eq.20}
\end{equation}
where $\lambda_r$ is the resonance wavelength (680~nm in the experiment) and $n_{\text{eff}}$ is the effective refractive index of the WGM mode. Rearranging Eq. ~\eqref{eq:Eq.20} to solve for $\Delta n$, we get $\Delta n \approx \left( \Delta \lambda / \lambda_r \right) \cdot n_{\text{eff}}$. Given that $\lambda_r = 680$~nm and assuming $n_{\text{eff}} = 2.5$, we estimate $\Delta n = \left( 50 \times 10^{-15}~\text{m} / 680 \times 10^{-9}~\text{m} \right) \cdot 2.5 \approx 1.84 \times 10^{-7}$. Therefore, the estimated change in the effective refractive index per single-molecule binding event is approximately $1 \times 10^{-7}$~RIU. This is a rough estimation, but it gives the correct order of magnitude for $\Delta n$ achieved in the benchmark paper. This result has two important implications:  
1) Achieving single-molecule detection is possible with a system that has a limit of detection (LOD) of at least $\sim 10^{-7}$~RIU;  
2) Our proposed approach already achieves an LOD of $\sim 10^{-7}$~RIU with photonic cavities having $Q = 10^5$ (see Table~\ref{tab:Tab2}), which is at least three orders of magnitude lower in $Q$ compared to the $Q > 10^8$ used in the benchmark study.

In addition, it is worth comparing the proposed approach with a commercially available alternative. Commercially available platforms such as Delta Life Science’s \textit{inQuiQ} system~\cite{delta2025} utilize microring resonator (MRR) structures to detect analyte-induced refractive index changes via resonance wavelength shifts, achieving state-of-the-art sensitivity with a baseline noise level of $\sim$0.01 RU (corresponding to $\sim 1 \times 10^{-8}$~RIU) at a 1~Hz readout (for ease of comparison, in some cases the calculated refractive index change (in RIU) is converted to resonance units (RU) --- a standard metric in SPR and other sensing techniques --- using the relation $1\,\mathrm{RU} = 10^{-6}\,\mathrm{RIU}$). According to publicly available information, this sensitivity is achieved using arrays of moderately high-$Q$ MRRs ($Q \approx 10^5$), operated near critical coupling to maximize the resonance shift per refractive index change $\Delta n$, and supported by active temperature stabilization and signal averaging to suppress baseline drift and noise. While highly effective, this approach relies on precise spectral tracking, which imposes integration challenges and demands high-resolution optical readout. It is unlikely that such systems employ on-chip integrated spectrometers, as achieving sub-picometer resolution with current CMOS-compatible technologies remains unfeasible; recent integrated spectrometer designs offer resolutions on the order of $\sim$100~pm~\cite{agneter2024}, which is several orders of magnitude too coarse to resolve the required spectral shifts. In contrast, our platform employs a fundamentally different transduction mechanism based on TCSPC to monitor lifetime shifts of embedded quantum emitters in silicon, directly modulated by changes in the LDOS. Importantly, this lifetime-based approach enables comparable sensitivity with photonic cavities having the same $Q$ values ($10^5$–$10^6$), significantly relaxing fabrication requirements. This makes our approach more scalable and robust, with inherent advantages in CMOS compatibility, noise resilience, and full on-chip integration without the need for spectral scanning or external tunable sources.

Given this, measuring the quantum emitter emission lifetime (via TCSPC) instead of tracking resonance wavelength shifts (via optical spectroscopy) offers several distinct practical and fundamental advantages:
\begin{enumerate}
	\item \textbf{Intrinsic sensitivity to LDOS changes:} Lifetime directly reflects variations in the local density of optical states (LDOS). Even subtle environmental modifications can induce measurable lifetime shifts, potentially providing greater sensitivity than wavelength shifts --- particularly for detecting very small refractive index changes.
	\item \textbf{Reduced complexity (no spectral measurements):} Lifetime measurements are performed at a single wavelength and do not require wavelength-resolved spectroscopy. This enables simpler and more compact integration with on-chip photodetectors (e.g., SPADs), which are significantly cheaper and less complex than high-resolution optical spectrometers. Notably, achieving a resolution of 50~fm with on-chip integrated spectrometers remains unfeasible with current technology. For instance, one of the latest CMOS optoelectronic spectrometers based on photonic integrated circuits demonstrated a resolution of only $\sim$93~pm --- approximately three orders of magnitude lower than what would be required for precise resonance tracking~\cite{agneter2024}.
	\item \textbf{Lower susceptibility to temperature fluctuations:} Emission lifetime is generally less affected by temperature-induced shifts compared to resonance wavelength, offering improved thermal stability and measurement reproducibility.
\end{enumerate}

\section{Physical basis of the signal transduction amplifier\label{Signal amplifier}}

The above discussions indicate that this approach may be suitable for use without any effective signal transduction amplifier of the local refractive index change signal (such as an aptamer-functionalized hydrogel (aptagel) layer~\cite{park2025}, by which the interaction areas schematically depicted in Figure~\ref{fig:figure1} can be covered), as it offers LODs on the order of $10^{-7}$–$10^{-9}$~RIU, which is already potentially sufficient for single-molecule detection. However, it is nonetheless interesting to analyze the role of such a mechanical-optical amplifier and to visualize its potential impact on sensing performance. 

For clarity, we note here that an aptagel is a hydrogel functionalized with specific aptamers (short single-stranded DNA or RNA sequences) that selectively bind target molecules. Upon binding, the aptamer undergoes a conformational change, which can trigger a macroscopic collapse or shrinkage of the hydrogel network (aptagel collapse). This collapse increases the local polymer density (polymer volume fraction), thereby increasing the local refractive index. In this way, the aptagel acts as a signal amplifier: a single molecular binding event can produce a much larger effective $\Delta n$ than direct binding of the molecule to the cavity surface. This differs from multiple identical molecules binding simultaneously, because here the amplification is not merely additive but comes from a cooperative volumetric change in the material.

Eq.~\eqref{eq:Eq.13} can be rearranged as follows:
\begin{equation}
	\frac{\Delta n_{\text{opt}}}{\Delta \tau / \tau_0} = \frac{n_{\text{eff}}}{2Q}.
	\label{eq:Eq.21}
\end{equation}
Eq.~\eqref{eq:Eq.21} expresses that, for a given cavity (i.e., fixed $Q$ and $n_{\text{eff}}$), the relationship between refractive index changes $\Delta n$ and the measurable lifetime shift $\Delta \tau / \tau_0$ is fixed at the optimal operating point. Therefore, if one increases the actual refractive index shift $\Delta n$ --- for example, due to aptagel collapse upon analyte binding — then the same sensing response (i.e., the same slope sensitivity) can be achieved with a higher $\Delta \tau / \tau_0$. In other words, while the aptagel does not alter the fundamental sensitivity of the photonic cavity (as governed by $Q$ and $n_{\text{eff}}$), it significantly increases the local refractive index change $\Delta n$ upon analyte binding. This, in turn, enables a proportional increase in the required measurable lifetime change $\Delta \tau / \tau_0$, thereby relaxing the timing resolution demands of the TCSPC detection system. 

To illustrate this concept, consider the following example: Without an aptagel, the binding of an analyte may induce a small refractive index change of approximately $1.2 \times 10^{-7}$~RIU. For a photonic cavity with a quality factor $Q = 10^6$, this corresponds to a relative lifetime change $\Delta \tau / \tau_0$ of about 0.3\%. Detecting such a small change would require a time resolution of around 60~ps (see Table~\ref{tab:Tab3}) — a capability available in high-end TCSPC systems. In contrast, when a responsive aptagel is employed, the same analyte binding event could lead to a significantly larger refractive index change — for example, $1.2 \times 10^{-6}$~RIU. This results in a relative lifetime change of approximately 3\%, which would translate to a TCSPC absolute resolution requirement of $\sim$600~ps. Such changes are much more easily detectable and can be resolved using standard, lower-cost TCSPC instrumentation. In this way, the aptagel does not increase the intrinsic sensitivity of the photonic cavity but instead relaxes the detection requirements by amplifying the local refractive index change resulting from analyte binding.

In addition, based on some realistic assumptions, we have attempted to visualize the effect of the aptagel in comparison to direct molecular binding to the sensor surface. When a single molecule binds directly to the surface of a SiO$_2$ cavity, it perturbs the cavity's optical mode through its local refractive index contribution. This perturbation is relatively small, as the molecule interacts only with the evanescent tail of the cavity mode. Let us assume the following: the refractive index of the molecule is $n_{\text{mol}} \approx 1.45$, the background medium has a refractive index of $n_{\text{bg}} \approx 1.33$, the volume of a single molecule is $V_{\text{mol}} \approx 10^2$~nm$^3$, and the mode volume of the cavity is $V_{\text{mode}} \approx 10^5$~nm$^3$. Then, the effective refractive index perturbation can be estimated as $\Delta n^{\text{direct}} = (n_{\text{mol}} - n_{\text{bg}}) \cdot \left( V_{\text{mol}} / V_{\text{mode}} \right) \cdot f_{\text{field}}(\vec{r})$, where $f_{\text{field}}(\vec{r})$ is the normalized field intensity at the molecule’s position $\vec{r}$, relative to the mode field maximum. Assuming $f_{\text{field}}(\vec{r}) = 10^{-3}$, the resulting $\Delta n^{\text{direct}} \approx 1.2 \times 10^{-7}$~RIU, which is in good agreement with our earlier estimation based on Ref.~\cite{armani2007}.

\begin{figure}[ht]
	\centering
	\includegraphics[width=\linewidth]{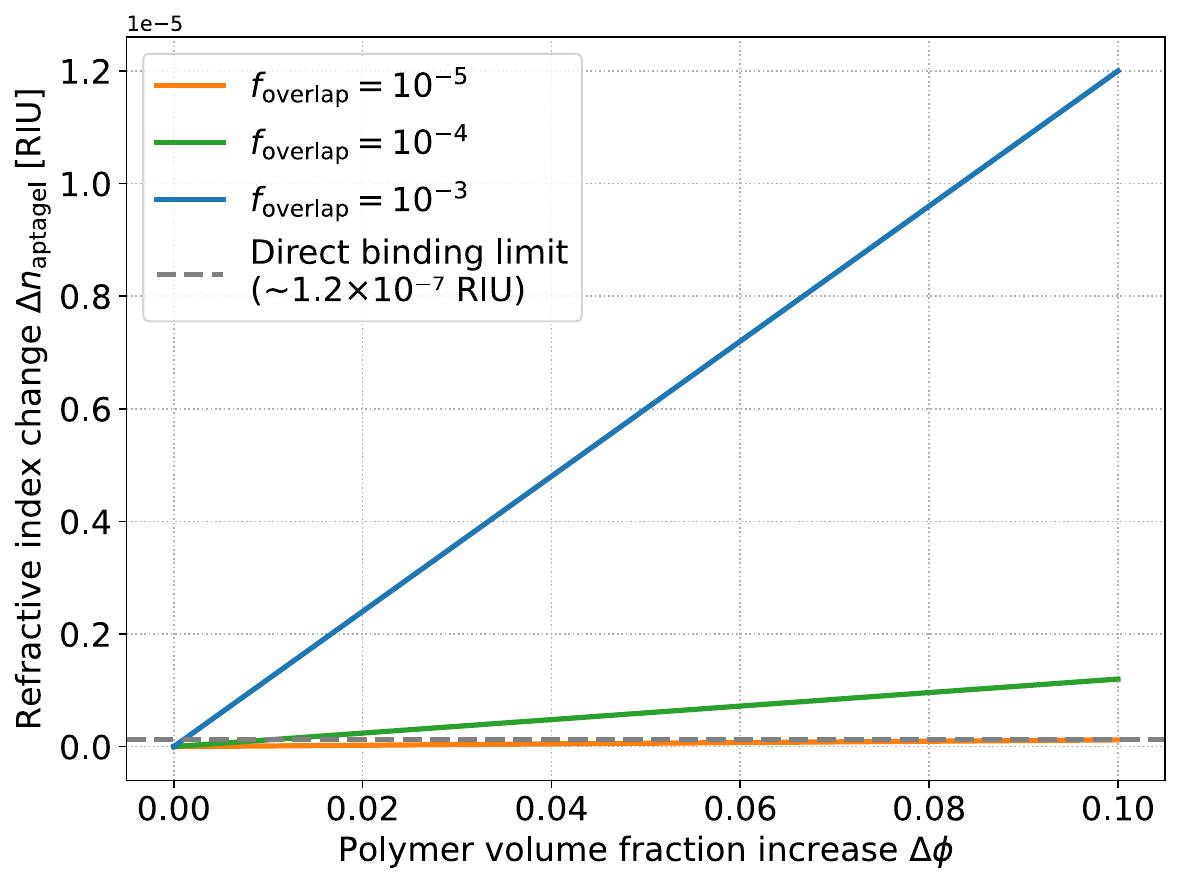}
	\caption{Dependence of $\Delta n^{\text{aptagel}}$ on the polymer volume fraction increase $\Delta \phi$ for different $f_{\text{overlap}}$ values.}
	\label{fig:figure6}
\end{figure}

In the case of aptagel-based amplification, the effective refractive index perturbation can be estimated as:
\begin{equation}
	\Delta n^{\text{aptagel}} = \Delta \phi \cdot (n_{\text{gel}} - n_{\text{water}}) \cdot f_{\text{overlap}},
	\label{eq:Eq.22}
\end{equation}
where $\Delta \phi$ is the increase in polymer volume fraction due to aptagel collapse (typically in the range 0.01–0.1), and $f_{\text{overlap}}$ is the field-weighted overlap between the collapsed gel and the cavity mode. The refractive indices of the gel and water are assumed to be $n_{\text{gel}} \approx 1.45$ and $n_{\text{water}} \approx 1.33$, respectively. Figure~\ref{fig:figure6} shows the dependence of $\Delta n^{\text{aptagel}}$ on the polymer volume fraction increase $\Delta \phi$ for different values of $f_{\text{overlap}}$.

At $\Delta \phi \rightarrow 0$, the curves closely approach the direct binding limit (dashed grey line), indicating that the aptagel has no amplification effect and behaves similarly to a single molecule bound to the surface. As $\Delta \phi$ increases, and for sufficiently high $f_{\text{overlap}}$ values ($10^{-3}$ or $10^{-4}$), even modest gel collapse results in a substantial increase in $\Delta n$, pushing the response well above the detection threshold and significantly facilitating measurement. This highlights the core advantage of aptagels — or any other amplification-capable functional layers: they can transform weak molecular binding signals into robust, volumetric refractive index changes.

\section{Discussion and Conclusion\label{Conclusion}}
We have presented a comprehensive theoretical framework for a new mechanism of probing refractive index changes based on Purcell-enhanced lifetime modulation of quantum emitters embedded in silicon photonic cavities. By operating the system off-resonance --- at the point of maximum slope in the Purcell response --- we identify a regime in which small refractive index perturbations produce large, linear, and readily measurable changes in fluorescence lifetime. This off-resonance strategy not only unlocks the full sensitivity potential of the system but also significantly relaxes the time-resolution requirements of the TCSPC instrumentation.

Our analysis demonstrates that, for quality factors in the range of $Q = 10^5$–$10^7$, the proposed sensing scheme achieves refractive index detection limits down to $10^{-9}$~RIU, with linear dynamic ranges suitable for single-molecule detection. Compared to conventional resonance-shift sensors, this lifetime-based approach offers several practical advantages: it is less susceptible to thermal and spectral noise, does not require high-resolution optical spectroscopy, and is fully compatible with integrated silicon photonic platforms. The use of long-lived emitters, such as T-centers, further enhances the system’s practicality by enabling high-$Q$ sensitivity to be accessed using standard, cost-effective TCSPC detectors.

\begin{figure*}[ht]
	\centering
	\includegraphics[width=\linewidth]{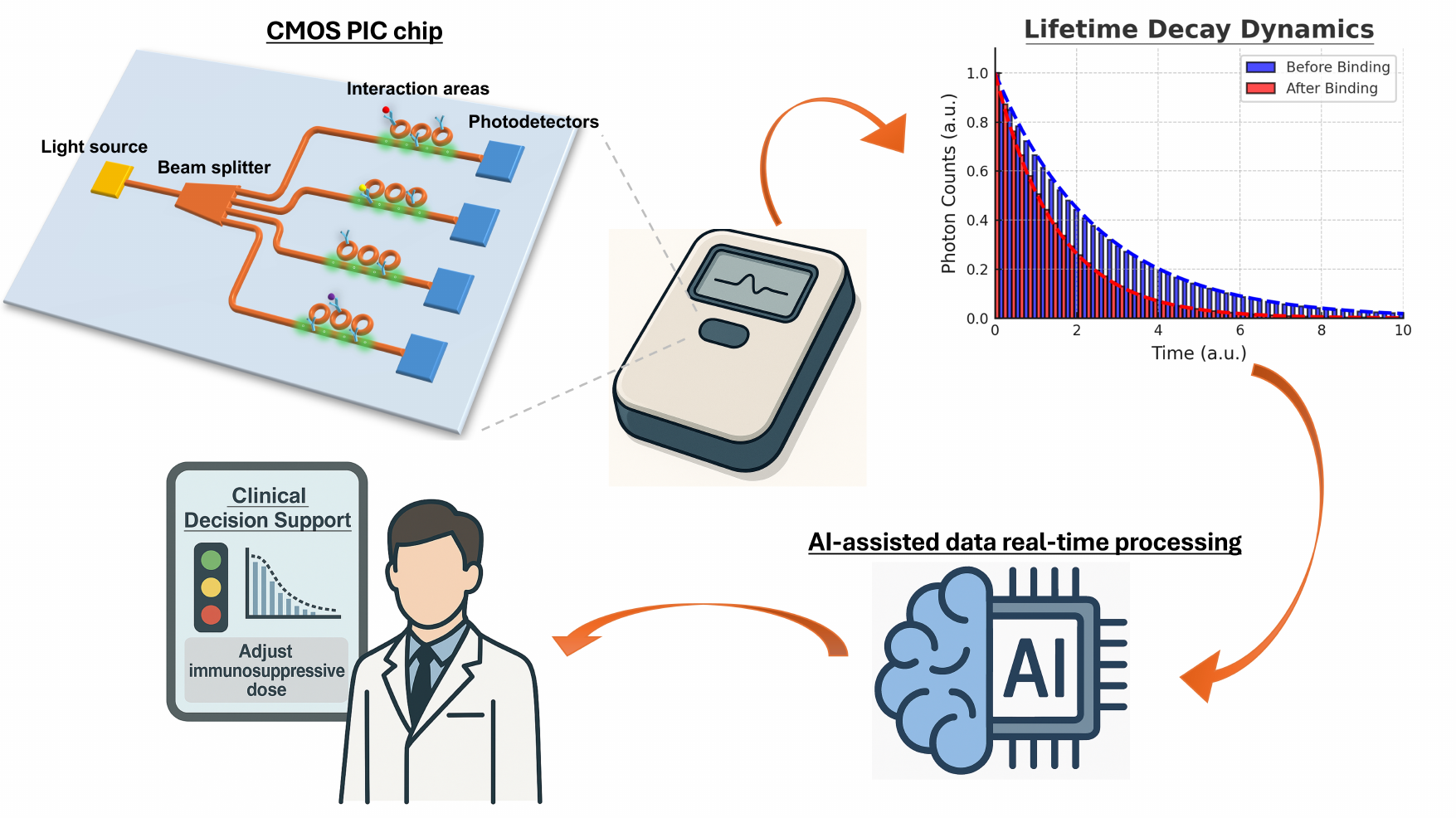}
	\caption{Possible application pathway of the proposed sensing platform for single-molecule interaction studies in personalized medicine.}
	\label{fig:figure7}
\end{figure*}

In terms of practical implementation, the sensing concept is directly compatible with silicon chips integrating single-photon avalanche diodes (SPADs), enabling label-free, on-chip lifetime measurements with high timing precision. Quality factor stability and cavity alignment --- often critical challenges in high-$Q$ resonance-shift sensors --- are less stringent in our approach due to the off-resonance operating regime, which maintains sensitivity even in the presence of moderate cavity detuning. According to Eq.~\eqref{eq:Eq.12} and as shown in Table~\ref{tab:Tab1}, the sensitivity of the proposed lifetime-based sensing scheme scales linearly with the cavity quality factor $Q$. Consequently, any relative variation in $Q$ directly translates into an equivalent relative change in sensitivity; for example, a $\pm 5\%$ fluctuation in $Q$ will result in approximately a $\pm 5\%$ change in sensitivity. This relationship underscores the importance of maintaining $Q$-factor stability in practical implementations, particularly when targeting ultra--low detection limits. Moreover, multiplexed layouts (Fig.~\ref{fig:figure1}) can allow parallel operation across multiple sensing channels, increasing throughput and enabling multi-analyte detection.

Figure~\ref{fig:figure7} outlines a possible application pathway in which the proposed platform serves as a basis for single-molecule interaction studies in personalized medicine. Here, a CMOS photonic integrated circuit (PIC) with integrated SPADs and interaction areas performs lifetime-based sensing. Real-time decay dynamics are processed via AI-assisted algorithms to extract binding events, and the resulting information can directly feed into clinical decision support systems, for example, to optimize immunosuppressive drug dosing in transplant patients.

While full TCSPC decay acquisition can take longer than the timescales of some single-molecule binding/unbinding events, detection speed can be improved by using high-brightness emitters, optimized photon extraction, and fast analysis techniques (e.g., phasor plots, running-window fits, or machine-learning-based event detection) that operate on partial decay data. Such approaches enable practical lifetime-based sensing even in dynamic environments and, while may not achieve instantaneous detection like resonance shift sensors, it can still operate effectively within the binding/unbinding timescales relevant to many biosensing applications. Furthermore, employing a pulsed laser with a MHz repetition rate together with high-speed acquisition and post-processing hardware ensures that TCSPC does not present a limitation for the proposed detection scheme.

Additionally, incorporating functional layers such as aptagels can act as signal transduction amplifiers, further increasing the effective refractive index shift and thereby reducing the demands on TCSPC detection performance.

In summary, this work establishes a fundamentally different and technologically viable sensing modality --- one that combines the robustness of quantum emitters with the practicality of lifetime-based detection, compatibility with SPAD arrays for label-free operation, and seamless integration into scalable PIC platforms. While silicon has been used here as a demonstrator owing to its CMOS compatibility and mature fabrication ecosystem, the generic nature of the proposed approach makes it readily adaptable to other PIC materials in which room-temperature operation of quantum emitters has already been experimentally demonstrated, such as diamond, silicon carbide, and silicon nitride. This versatility opens a clear pathway toward next-generation, on-chip sensors with quantum-enabled sensitivity, operable at room temperature and without reliance on spectral resolution.

\section*{Author Contributions}

Y.M. conceived the idea, performed the calculations, developed the Python scripts, and wrote the original draft of the manuscript. A.L. proposed the model of the averaged fluorescence decay (Eq.~\eqref{eq:Eq.18}), assisted with the calculations, and contributed to manuscript writing. All authors have read and approved the final version of the manuscript.

\begin{acknowledgments}
No external funding was received for this research. The authors would like to thank the Armed Forces of Ukraine for providing security to perform this work.
\end{acknowledgments}

\section*{Data Availability}
This study did not generate new experimental data. All Python scripts used for analytical calculations and figure generation, as well as a supporting PDF document detailing LDR calculation, are available at the OSF repository (under the Files tab): https://doi.org/10.17605/OSF.IO/QDYNX.

\bibliography{references}

\onecolumngrid
\section*{Annex: Linear Dynamic Range (LDR) Calculation}
The linear dynamic range (LDR) is defined as the range of refractive index changes $\Delta n$ around the point of maximal linearity ($n_0$) for which the relative lifetime response deviates less than $\pm\varepsilon$ from the linearized model (Eq.12 in the main text). This statement can be written as:
\begin{equation}
	\left| \frac{\tau(n)-\tau_0}{\tau_0} \right| \approx \left|\frac{2Q\Delta n}{n_\mathrm{eff}} \right|,
	\tag{S1} \label{eq:Eq.S01}
\end{equation}
or
\begin{equation}
	\left| \frac{\tau(n)-\tau_0}{\tau_0} - \frac{2Q\Delta n}{n_\mathrm{eff}} \right| < \varepsilon,
	\tag{S2}\label{eq:Eq.S02}
\end{equation}
where $\Delta n = n - n_\mathrm{0}$, i.e., deviation from the optimal off-resonance point. According to Eq.(7) of the main text: $\tau(n)=\tau_\mathrm{eff}(1+x^2)$, where $x = (2Q\Delta n/n_\mathrm{eff})$; if $x = 0$ (i.e. $\Delta n = 0$): $\tau(n) = \tau_\mathrm{eff}$; if $x = 1$ (i.e. $\Delta n = n_\mathrm{eff}/2Q$): $\tau(n) = \tau_0 = 2\tau_\mathrm{eff}$. Now we substitute these values into the left-hand side of Eq.~\eqref{eq:Eq.S01}:
\begin{equation}
	\left| \frac{\tau(n)-\tau_0}{\tau_0} \right| = \left| \frac{\tau(n)}{\tau_0} - 1 \right| = \left| \frac{\tau(n)}{2\tau_\mathrm{eff}} - 1 \right| = \left| \frac{1+x^2}{2} - 1 \right| = \left| \frac{x^2-1}{2} \right|.
	\tag{S3} \label{eq:Eq.S03}
\end{equation}
It should be noted that due to this division by $\tau_0$, $\tau_0$ is normalized to be 1. Finally, Eq.~\eqref{eq:Eq.S02} can be re-written as:
\begin{equation}
	\left| \frac{x^2-1}{2} - x \right| < \varepsilon.
	\tag{S4}\label{eq:Eq.S04}
\end{equation}
Eq.~\eqref{eq:Eq.S04} can be used for numerical LDR calculation as it is shown in the Python example in the end of this document.

Furthermore, assuming that 1) the quadradic function on the left-hand side of Eq.~\eqref{eq:Eq.S04} is centered at the linearization point, and 2) the domain of acceptable error ($x$ values satisfying the inequality) is symmetric around that point (i.e. $x = 1$), a closed-form analytical expression can be obtained for a quick estimate and comparison with numerical values. From the above considerations, the normalized lifetime response reads as:
\begin{equation}
	\frac{\tau(n)}{\tau_0} = \frac{1 + x^2}{2}.
	\tag{S5}\label{eq:Eq.S05}
\end{equation}
Near the point \( x = 1 \), we linearize \( \tau(n)/\tau_0 \) using a Taylor expansion:
\begin{equation}
	f(x) = \frac{1 + x^2}{2} \Rightarrow f'(x) = x.
	\tag{S6}\label{eq:Eq.S06}
\end{equation}
Expanding around \( x = 1 \):
\begin{equation}
	\frac{\tau(n)}{\tau_0} \approx f(1) + f'(1)(x - 1) = 1 + (x - 1) = x.
	\tag{S7}\label{eq:Eq.S07}
\end{equation}
Thus, the linearized model near \( x = 1 \) is simply:
\begin{equation}
	\frac{\tau(n)}{\tau_0} \approx x.
	\tag{S8}\label{eq:Eq.S08}
\end{equation}
To quantify the LDR, we define it as the range of \( x \) over which the deviation between the nonlinear model (Eq.~\eqref{eq:Eq.S05}) and its linear approximation (Eq.~\eqref{eq:Eq.S08}) remains equal to \( \varepsilon \):
\begin{equation}
	\left| \frac{1 + x^2}{2} - x \right| = \varepsilon.
	\tag{S9}\label{eq:Eq.S09}
\end{equation}
Rewriting Eq.~\eqref{eq:Eq.S09}:
\begin{equation}
	\left| \frac{(x - 1)^2}{2} \right| = \varepsilon \Rightarrow (x - 1)^2 = 2\varepsilon,
	\tag{S10}\label{eq:Eq.S10}
\end{equation}
where the positive root is taken because the function is concave and the approximation lies below the curve. Solving Eq.~\eqref{eq:Eq.S10} gives:
\begin{equation}
	x_{1,2} = 1 \pm \sqrt{2\varepsilon}.
	\tag{S11}\label{eq:Eq.S11}
\end{equation}
Now we convert back to \(\Delta n\). We recall that:
\begin{equation}
	x = \frac{2Q \Delta n}{n_\mathrm{eff}} \Rightarrow \Delta n = \frac{n_\mathrm{eff}}{2Q}x.
	\tag{S12}\label{eq:Eq.S12}
\end{equation}
Thus, the corresponding values of \( \Delta n \) are:
\begin{equation}
	\Delta n_1 = \frac{n_\mathrm{eff}}{2Q}(1 - \sqrt{2\varepsilon}), \quad
	\Delta n_2 = \frac{n_\mathrm{eff}}{2Q}(1 + \sqrt{2\varepsilon}).
	\tag{S13}\label{eq:Eq.S13}
\end{equation}
Final analytical expression for LDR width is therefore:
\begin{equation}
	\text{LDR width} = \Delta n_2 - \Delta n_1 = \frac{n_\mathrm{eff}}{Q} \sqrt{2\varepsilon}.
	\tag{S14}\label{eq:Eq.S14}
\end{equation}
Eq.~\eqref{eq:Eq.S14} provides a closed-form expression for fast analytical estimate for the LDR width. Table~\ref{tab:LDR_comparison} provides a comparison between analytical (Eq.~\eqref{eq:Eq.S14}) and numerical (Eq.~\eqref{eq:Eq.S04}) approaches.

\renewcommand{\thetable}{S\arabic{table}}

\begin{table}[h!]
	\centering
	\caption{LDR comparison}
	\begin{tabular}{|c|c|c|c|}
		\hline
		\textbf{Q} & \textbf{Numerical LDR} & \textbf{Analytical LDR} & \textbf{Relative difference} \\
		\hline
		$1\times10^5$ & 1.11820112e-05 & 1.11803399e-05 & 0.01\% \\
		$1\times10^6$ & 1.11820112e-06 & 1.11803399e-06 & 0.01\% \\
		$1\times10^7$ & 1.11820112e-07 & 1.11803399e-07 & 0.01\% \\
		\hline
	\end{tabular}
	\label{tab:LDR_comparison}
\end{table}

As it follows from Table~\ref{tab:LDR_comparison}, results obtained using analytical and numerical approaches are in good agreement. To summarize, Eq.~\eqref{eq:Eq.S04} provides a direct and intuitive way to evaluate the deviation between the nonlinear response and its linear approximation, making it ideal for empirical LDR determination. However, the expression provided by Eq.~\eqref{eq:Eq.S14} offers a closed-form estimate useful for quick comparisons and scaling analysis.

\section*{Python code}

\begin{lstlisting}[style=mypython, caption={Python code for computing LDR width}]
import numpy as np
	
def tau_over_tau0(n, neff, Q):
	# tau(n)/tau_0 as a function of n, with tau_0 = 2*tau_eff (i.e., tau(n_0))
	x = 2 * Q * (n - neff) / neff
	return (1 + x**2) / 2
	
def compute_LDR(Q, neff, deviation):
	n0 = neff + neff / (2 * Q)
	tau0 = 1.0
	slope = (2 * Q) / neff
	
	dn_range = 100 / Q
	dn_values = np.linspace(-dn_range, dn_range, 1_000_000)
	n_values = n0 + dn_values
	
	tau_norm = tau_over_tau0(n_values, neff, Q)
	delta_tau_norm = tau_norm - tau0
	linear = slope * (n_values - n0)
	
	# Compute deviation from linear approximation (Eq. 12)
	abs_dev = np.abs(delta_tau_norm - linear)
	
	# Symmetric expansion from n_0 to find the LDR width
	center = np.argmin(np.abs(dn_values))
	left, right = center, center
	
	while left > 0 and abs_dev[left] < deviation:
		left -= 1
	while right < len(abs_dev) - 1 and abs_dev[right] < deviation:
		right += 1
	
	ldr_width = dn_values[right] - dn_values[left]
	
	print(f"Q={Q:.0e}, LDR width = {ldr_width:.2e}, deviation threshold = {deviation}")
	return ldr_width
	
for Q in [1e5, 1e6, 1e7]:
	compute_LDR(Q, neff=2.5, deviation=0.10)
\end{lstlisting}

\end{document}